\documentclass[fdp,a4paper,fleqn]{w-art}
\usepackage{times,w-thm}
\usepackage{amssymb}
\theoremstyle{plain}

\theoremstyle{definition}

\usepackage[]{graphicx}
\begin{document}
\pagespan{1}{}



\title[Renormalizability  and Phenomenology of $\theta$-expanded NCGFT]
{Renormalizability  and Phenomenology of $\theta$-expanded Noncommutative Gauge Field Theory}


\author[Josip Trampeti\' c]{Josip Trampeti\' c\inst{1,}%
  \footnote{Corresponding author\quad E-mail:~\textsf{josipt@rex.irb.hr},
	   }}
\address[\inst{1}]{Theoretical Physics Division, Rudjer Bo\v skovi\' c Institute, Zagreb, Croatia}
\begin{abstract}
In this article we consider $\theta$-expanded noncommutative gauge field theory, 
constructed at the first order in noncommutative parameter $\theta$,
as an effective, anomaly free theory, with one-loop renormalizable gauge sector. 
Related phenomenology with emphasis on the standard model forbidden decays, is discussed.
Experimental possibilities of $Z\rightarrow \gamma\gamma$ decay are analyzed and a
firm bound to the scale of the noncommutativity parameter is set around few TeV's.
\end{abstract}
\maketitle                   


\section{ Introduction }     
One of the first example of noncommutativity (NC) is well known Heisenberg algebra.
Motivations to construct models on noncommutative space-time are comming from:  String Theory,
Quantum Gravity, Lorentz invariance breaking, and by its own right.
The star product definition is as usual. The $\star$-commutator  
and Moyal-Weyl $\star$-product of two functions are: 
\begin{eqnarray}   
[x^\mu\stackrel{\star}{,}x^\nu]  &=& x^\mu \star x^\nu - x^\nu \star x^\mu 
=ih\theta^{\mu\nu},
\;\;\;
(f \star g) (x) =  e^{-\frac{i}{2} \theta^{\mu\nu} 
\frac{\partial}{\partial{x^{\mu}}} \frac{\partial}{\partial{y^{\nu}}}}
f(x)g(y)|_{y \rightarrow x}\,.
\label{1}
\end{eqnarray} 
Noncommutative space is flat Minkowski space were commutative coordinates 
$x^{\mu}$ are replaced by the NC ones ${\hat x}^{\mu}$,
satisfying the same commutator as above; that is: 
\begin{equation}
x^{\mu}\rightarrow {\hat x}^{\mu}\, \Rightarrow
\left[{\hat x}^{\mu},{\hat x}^{\nu} \right]=ih\theta^{\mu\nu}, 
\quad \left[\theta^{\mu\nu},{\hat x}^{\rho} \right]=0.
\label{2}                
\end{equation} 
Here $\theta$ is constant, antisymmetric and real ${4\times4}$ matrix;  
${h=1/\Lambda^2_{\rm NC}}$ is noncommutative deformation parameter.
Symmetry in our model \cite{Wess}, using Seiberg-Witten map (SW) \cite{Seiberg:1999vs} 
is extended to enveloping algebra \cite{Wess,Calmet:2001na}.
Any enveloping algebra based model is essentialy double expansion
in power series in $\theta$ \cite{Wess,Calmet:2001na,Blazenka,Aschieri:2002mc,Goran}. 
In principle SW map express noncommutative functionals 
(parameters and functions of fields) 
spanned on the noncommutative space as a 
local functionals spanned on commutative space. 

To obtain the action we first do the Seiberg-Witten expansion of 
NC fields in terms of commutative ones and second we 
expand the $\star$-product. This procedure generets tower of new vertices, 
however it is valid for
any gauge group and arbitrary matter representation.
Also there is no charge quantization problem and no UV/IR mixing
\cite{UV/IR}.
Unitarity is satisfied for $\theta^{0i}=0$ and $\theta^{ij}\not=0$ 
\cite{Seiberg:2000gc,Gomis:2000zz}; however
careful canonical quantization produces always unitary theory. 
By covariant generalization of the condition $\theta^{0i}=0$ to: 
 \begin{eqnarray} 
\theta_{\mu\nu}\theta^{\mu\nu} &=&-\theta^2
=2h^2\left(\sum_{{i,j=1 \atop i<j}}^{3}({\theta}^{ij})^2-\sum_{i=1}^{3}({\theta}^{0i})^2\right)
= \frac{2}{\Lambda_{\rm NC}^4}
 \left (\vec{B}_{\theta}^2 - \vec{E}_{\theta}^2 \right ) >0\;, 
 \label{3}
\end{eqnarray}
which is known as {\it perturbative unitarity condition} \cite{Carroll:2001ws},
there is no difficulties with unitarity in NC gauge theories.
Finally covariant noncommutative Higgs and Yukawa couplings were possible to construct \cite{Blazenka}.

There are two essential points in which NC gauge field theory (NCGFT) differ from
standard model (SM) gauge theories. 
The breakdown of Lorentz invariance with
respect to a fixed nonzero background field $\theta^{\mu\nu}$
(which fixes preferred directions) 
and the appearance of new interactions 
and the modification of standard ones. 
For example, triple--neutral--gauge boson, 
two fermion--two gauge bosons, 
direct photon-neutrino couplings, etc.
Both properties have a common origin and appear in a number of phenomena
at very high energies and/or very short distances.

In this article we discuss $\theta$-expanded theories, constructed as an effective, 
anomaly free \cite{Brandt:2003fx} and one-loop renormalizable NCGFT 
\cite{Bichl:2001cq,Buric:2005xe,Buric:2006wm,Latas:2007eu,Martin:2006gw},
at the first order in noncommutative parameter $\theta$. 
We also consider the $4\psi$ divergences in noncommutative chiral models for fermions;
specifically we discuss the U(1) and the SU(2) cases \cite{Buric:2007ix}.
Finally we discuss related phenomenology and determine 
the scale of noncommutativity $\Lambda_{\rm NC}$, \cite{Buric:2007qx,Josip,Melic:2005hb}.


\section{Properties of $\theta$-expanded noncommutative gauge field theory }
(a) Noncommutative gauge transformation: \\
Consider infinitesimal noncommutative local gauge 
transformation $\hat\delta$
of a fundamental matter field that carries a representation $\rho_\Psi$,
which is in Abelian case fixed by the hypercharge,
\begin{eqnarray} 
\hat\delta \widehat\Psi = i \rho_{\Psi}(\widehat\Lambda) \star \widehat\Psi \;.
\label{4}
\end{eqnarray}
(b) Covariant coordinates: 
\begin{eqnarray} 
{\hat x}^\mu = x^\mu + h\theta^{\mu\nu} {\widehat A}_\nu 
\label{5}
\end{eqnarray}
were in noncommutative theory introduced in analogy to covariant derivatives in ordinary theory.\\
(c) Locality of the theory:\\
A $\star$-product of two ordinary functions $f(x)$ and $g(x)$,
determined by a Poisson tensor $\theta^{\mu\nu}$ and written in the form of expansion, 
is local function of 
$f$ and $g$ with finite number of derivatives at each order in $\theta$:
\begin{eqnarray} 
f(x) \star g(x) = f(x) \cdot g(x) + \frac{i}{2}\theta^{\mu\nu} \partial_\mu f(x) 
\cdot \partial_\nu g(x) + \mathcal{O}(\theta^2)\,.
\label{6}
\end{eqnarray}
(d) Gauge equivalence, and consistency conditions for the theory:\\	
Ordinary gauge
transformations $\delta A_\mu = \partial_\mu \Lambda + i[\Lambda,A_\mu]$
and $ \delta \Psi = i \Lambda\cdot \Psi$ 
induce noncommutative gauge transformations of the fields 
$\widehat A$, $\widehat \Psi$ with gauge parameter $\widehat \Lambda$ 
\begin{eqnarray} 
\delta \widehat A_\mu = \hat\delta \widehat A_\mu
\qquad \delta \widehat \Psi = \hat\delta \widehat \Psi\,.
\label{7}
\end{eqnarray}
Consistency require that any pair of noncommutative 
gauge parameters 
$\widehat\Lambda$, $\widehat{\Lambda'}$ satisfy
\begin{eqnarray} 
[\widehat\Lambda,\widehat{\Lambda'}] + i \delta_\Lambda \widehat{\Lambda'}
- i \delta_{\Lambda'} \widehat\Lambda = \widehat{[\Lambda,{\Lambda'}]}\,.
\label{8}
\end{eqnarray}
(e) Enveloping algebra-valued noncommutative gauge parameters and fields:\\
For the enveloping algebra-valued gauge transformation, the commutator 
\begin{eqnarray}
[\widehat\Lambda,\widehat\Lambda'] 
=  \frac{1}{2}\{\Lambda_a(x)\stackrel{\star}{,} \Lambda'_b(x)\}[T^a,T^b] 
+ \frac{1}{2} [\Lambda_a(x)\stackrel{\star}{,}\Lambda'_b(x)]\{T^a,T^b\}
\label{9}
\end{eqnarray}
of two Lie algebra-valued noncommutative gauge parameters 
$\widehat\Lambda = \Lambda_a(x) T^a$ and $\widehat\Lambda' = \Lambda'_a(x) T^a$
does not close in Lie. For noncommutative SU(N) the Lie algebra 
traceless condition is incompatible with commutator. 
So, for noncommutative gauge transformation we have extension to 
the enveloping algebra-valued gauge transformation expressed by 
the following expansion: 
\begin{eqnarray}
\widehat\Lambda = \Lambda^0_a(x) T^a +  \Lambda^1_{ab}(x) :T^a T^b:
+ \Lambda^2_{abc}(x) :T^a T^b T^c: + \ldots
\label{10}
\end{eqnarray}
(f) Seiberg-Witten map:\\
Closing condition for gauge transformation algebra are homogenus differential equations,
which are solved by iteration, order by order in noncommutative parameter $\theta$. 
Solutions are known as Seiberg-Witten map. Hermicity condition for the fields, 
up to the first order in Seiberg-Witten expansion, gives for gauge parameter, fermion and gauge fields
the following expressions:  
\begin{eqnarray} 
\widehat \Lambda &=& \Lambda +
\frac{h}{4} \theta^{\mu \nu}
           \{ V_{\nu}, \partial_{\mu}\Lambda\}+...
\;\;\;\;\;
\widehat\psi =  \psi - \, \frac{h}{2}\,\theta^{\alpha \beta} \,
\,\Big( V_{\alpha}  \,\partial_{\beta}
          - \frac{\mathrm{i}}{4}  \,
              [V_{\alpha}, V_{\beta}] \Big) \,\psi+...	      
\nonumber\\
\widehat V_\mu  &=& V_{\mu} +
    \frac{h}{4} \theta^{\alpha \beta}
           \{ \partial_{\alpha}V_{\mu} + F_{\alpha \mu},V_{\beta}\}+...
\label{11}	  \\
{\widehat F}_{\mu\nu} &=& \partial_{\mu} {\widehat V}_{\nu} - \partial_{\nu} {\widehat V}_{\mu}
- \mathrm{i}[{\widehat V}_{\mu}\stackrel{\star}{,}{\widehat V}_{\nu}]
 =F_{\mu\nu} +\frac{h}{4}\theta^{\rho\sigma}
\Big(2\{F_{\rho\mu},F_{\sigma\nu}\} -
\{ V_{\rho} ,(\partial_{\sigma} +D_{\sigma} )F_{\mu\nu} \}\Big)\,.
\nonumber
\end{eqnarray}

\section{Noncommutative gauge field theory framework proposal} 
Commutative GFT, that are renormalizable
 with minimal coupling, are extended 
in the same minimal fashion 
to the NC space with deformed gauge transformations. 
These deformations are not unique. For instance deformed action ${S_g}$
depends on the choice of representation. 
This derives from the fact that ${\widehat F}^{\mu\nu}$ is enveloping algebra
not Lee algebra valued. So called
`minimally coupled NC' gauge-invariant action is:
\begin{eqnarray}
S_{\rm NC}&=& S_g+S_{\psi}
=-\frac{1}{2}\;{\rm Tr}\;\int d^4x\; 
{\widehat F}_{\mu\nu} \star {\widehat F}^{\mu\nu}
+\mathrm{i}\;\int \mathrm{d}^4x\; \widehat{\bar \varphi}
\star \bar\sigma^\mu (\partial_\mu +\mathrm{i}\widehat A_\mu ) 
\star \widehat\varphi\,.
\label{12} 
\end{eqnarray}
The trace Tr in ${S_g}$ is over all representations.
${\widehat{\varphi}}$'s are the noncommutative Weyl spinors.
Applying Seiberg-Witten map on the above action up to first order in $\theta$
we obtain `minimal' actions
\begin{eqnarray}
S_g
&=& 
- \frac{1}{2} {\rm Tr} \int d^4 x \;
  F_{\mu \nu} F^{\mu \nu}
+h\,\theta^{\rho \sigma} {\rm Tr} \int d^4 x \;
\left[ \left( \frac{1}{4} F_{\rho \sigma} F_{\mu \nu}
-F_{\rho \mu} F_{\sigma \nu} \right) F^{\mu \nu} \right]\,,
\nonumber \\
S_{\psi}
&=&
\mathrm{i}\;\int d^4 x \;\bar\varphi\sigma^\mu (\partial_\mu
+\mathrm{i}  A_\mu )\varphi
-\frac{h}{8} \theta^{\mu\nu}
\Delta^{\alpha\beta\gamma}_{\mu\nu\rho}
\;\int d^4 x \;F_{\alpha\beta} \;\bar\varphi
\;\bar\sigma^\rho(\partial_\gamma
+\mathrm{i} A_\gamma)\varphi\,,
\nonumber\\
\Delta^{\alpha\beta\gamma}_{\mu\nu\rho}
&=& \varepsilon^{\alpha\beta\gamma\lambda}\varepsilon_{\lambda\mu\nu\rho}\,.
\label{13}
\end{eqnarray}

Clearly we do not know the meaning of `minimal coupling
concept' for some NCGFT in the NC space.  
However, renormalization is the principle that help us to find such 
acceptable couplings.
We learned that the renormalizability condition of some specific NCGFT 
requires introduction of the higher order noncommutative gauge interaction by
expanding general NC action in terms of NC field strengths.
This of course extends `NC minimal coupling' of 
the gauge action ${S_g}$ in (\ref{12}) to higher order:
\begin{equation}
S_g=-\frac{1}{2}\;{\rm Tr} \int \mathrm{d}^4x\,
\left(1-\frac{a-1}{2}h\theta_{\rho\sigma} \star \widehat F^{\rho\sigma} \right)
\star \widehat F_{\mu\nu} \star \widehat F^{\mu\nu} \,,
\label{14}
\end{equation}
with $a$ being free parameter determining renormalizable deformation.
This was possible due to the symmetry property of an object
$\theta_{\rho\sigma} \star \widehat F^{\rho\sigma}$.
SW map for NC field strength up to the first order in ${h\theta^{\mu\nu}}$ than gives:
\begin{eqnarray}
S_g &=&{\rm Tr} \int \mathrm{d}^4x\,\Big[-\frac{1}{2}
F_{\mu\nu}F^{\mu\nu}
+h\theta^{\mu\nu}\, \left(
\frac{a}{4} F_{\mu\nu}F_{\rho\sigma}-F_{\mu\rho}F_{\nu\sigma} \right)F^{\rho\sigma}\Big]\,.
\label{15}
\end{eqnarray}
In the chiral fermion sector the choice of Majorana spinors for the U(1) case gives
\begin{eqnarray}
S_{\psi}
&=&
\frac{\mathrm{i}}{2}\;\int d^4 x 
\Big[ \;\bar\psi\gamma^\mu (\partial_\mu
-\mathrm{i} \gamma_5 A_\mu )\psi
+\mathrm{i}\frac{h}{8} \theta^{\mu\nu}
\Delta^{\alpha\beta\gamma}_{\mu\nu\rho}
F_{\alpha\beta} \;\bar\psi
\;\gamma^\rho(\partial_\gamma
-\mathrm{i}\gamma_5 A_\gamma)\psi\,\Big]\,.
\label{16}
\end{eqnarray}
For the SU(2) case relevant expressions are given in \cite{Buric:2007ix}.

Proposed framework gives starting action
for the gauge and fermion sectors.
Requirement of renormalizability
fixes the freedom parameter $a$. That is, the
principle of renormalization determines NC renormalizable deformation.
Trace of three generators in the above action lead to  
dependence of the gauge group representation and
the choice of the trace corresponds to the choice of the group representation.

\section{Gauge sector}
\subsection{\bf Gauge sector of minimal NCSM}
Choosing vector field in the adjoint representation, i.e.  using a sum of three
traces over the standard model gauge group we have the following action
\begin{equation}
S^{\mbox{\tiny mNCSM}}_g = -\frac{1}{2} \int
d^4x \left( \frac{1}{{g'}^2} \mbox{Tr}_{\bf 1} + \frac{1}{g^2}
\mbox{Tr}_{\bf 2} +\frac{1}{g_s^2} \mbox{Tr}_{\bf 3}
\right) \,
\left(1-\frac{a-1}{2}h\theta_{\rho\sigma} \star \widehat F^{\rho\sigma} \right)
\star 
\widehat F_{\mu \nu}\star \widehat F^{\mu \nu} \, .
\label{17}
\end{equation}
In definition of $\mbox{Tr}_{\bf 1}$ we use usual representation of the hypercharge
$Y = \frac{1}{2} \left(\begin{array}{rr} 1 & 0 \\ 0 & -1\end{array}\right)$\,.
For the fundamental representations of SU(2) and SU(3) we have the generators
in $\mbox{Tr}_{\bf 2}$ and $\mbox{Tr}_{\bf 3}$, respectively.
In terms of physical fields, the gauge action then reads
\begin{eqnarray}
S^{\mbox{\tiny mNCSM}}_g &=&
-\frac{1}{2} \int d^4x \Big[\left( \frac{1}{2} {\cal A}_{\mu\nu}{\cal A}^{\mu\nu}
+ \, \mbox{Tr}\, {\cal B}_{\mu\nu} {\cal B}^{\mu\nu}
+ \, \mbox{Tr}\, G_{\mu\nu} G^{\mu\nu} \right)
\nonumber
\\ & & 
- \frac{1}{2} \, g_s\, {d^{abc}} \, h\theta^{\rho\sigma}
\left(
\frac{ a}{4}G^a_{\rho\sigma} G^b_{\mu\nu}
-G^a_{\rho\mu} G^b_{\sigma\nu}
\right)G^{\mu\nu,c}\Big]  \, ,
\label{18}
\end{eqnarray}
where $d^{abc}$ are totally symmetric SU(3) group coefficients
which come from the trace in (\ref{17}). 
The ${\cal A}_{\mu\nu}$, ${\cal B}_{\mu\nu}(=B_{\mu\nu}^aT_L^a)$
and $G_{\mu\nu}(=G_{\mu\nu}^aT_S^a)$ denote the U(1), 
$\rm SU(2)_L$ and $\rm SU(3)_c$ field strengths,
respectively:
\begin{eqnarray}
{\cal A}_{\mu\nu}&=&\partial_{\mu}{\cal A}_{\nu}-\partial_{\nu}{\cal A}_{\mu}\:,
\;\;\;
{\cal B}_{\mu\nu}^a = \partial_{\mu}B^a_{\nu}-\partial_{\nu}B^a_{\mu}
+g\;\epsilon^{abc}B^b_{\mu} B^c_{\nu}\:,
\nonumber\\
G_{\mu\nu}^a &=& \partial_{\mu}G^a_{\nu}-\partial_{\nu}G^a_{\mu}
+g_s\;f^{abc}G^b_{\mu} G^c_{\nu}\:.
\label{19}
\end{eqnarray}
For adjoint representation their is no new neutral electroweak triple gauge boson interactions.

\subsection{\bf Gauge sector of nonminimal NCSM}
The nmNCSM gauge sector action is given by Eq.(\ref{15})
where Tr is
trace over all massive particle multiplets with different quantum numbers 
in the model that have covariant derivative acting on them;
five multiplets for each generation of fermions and one Higgs multiplet.
Here $F_{\mu \nu}=\partial_{\mu} V_{\nu}-\partial_{\nu} V_{\mu}- \mathrm{i} [V_{\mu},V_{\nu}]$ is
standard model field strength, i.e. $V_{\mu}$ is the standard model gauge potential:
\begin{eqnarray} 
V^{\mu}&=&g'{\cal A}^{\mu}(x)Y + g\sum^3_{a=1}B^{\mu}_a(x)T^a_L 
+ g_s\sum^8_{b=1}G^{\mu}_{b}(x)T^b_S\,.
\label{20}
\end{eqnarray}
Matching the standard model action at zeroth order in $\theta$, three consistency conditions are
imposed producing final expression for triple gauge boson (TGB) action. In terms of 
the U(1), SU(2) and SU(3) field strengths, $f_{\mu \nu},\;\;B^a_{\mu \nu}$ and $G^b_{\mu \nu}$,
respectively, we have the following action
\begin{eqnarray} 
 S^{\rm nmNCSM}_{gauge}&=&S_{cl} =
-\frac{1}{4}\int d^4x f_{\mu \nu} f^{\mu \nu}
-\frac{1}{2}\int \hspace{-1mm}d^4x\, {\rm Tr}\left( B_{\mu \nu} B^{\mu \nu}\right)
-\frac{1}{2}
\int\hspace{-1mm} d^4x\, {\rm Tr}\left( G_{\mu \nu} G^{\mu \nu}\right)
\nonumber  \\&+&
{g'}^2\kappa_1 h\theta^{\rho\tau}\hspace{-2mm}\int \hspace{-1mm}d^4x\,
\left(\frac{a}{4}f_{\rho\tau}f_{\mu\nu}-f_{\mu\rho}f_{\nu\tau}\right)f^{\mu\nu}
 \nonumber \\
&+&g'g^2\kappa_2 \, h\theta^{\rho\tau}\hspace{-2mm}\int
\hspace{-1mm} d^4x \sum_{a=1}^{3}
\left[(\frac{ a}{4}f_{\rho\tau}B^a_{\mu\nu}-
f_{\mu\rho}B^a_{\nu\tau})B^{\mu\nu,a}\!+c.p.\right]
 \nonumber \\
&+&g'g^2_s\kappa_3\, h\theta^{\rho\tau}\hspace{-2mm}\int
\hspace{-1mm} d^4x \sum_{b=1}^{8}
\left[(\frac{a}{4}f_{\rho\tau}G^b_{\mu\nu}-
f_{\mu\rho}G^b_{\nu\tau})G^{\mu\nu,b}\!+c.p.\right] \,.
\label{21}
\end{eqnarray}
Three consistency conditions together with definitions of three couplings 
$\kappa_i$ and the requirement that 
$1/g^2_i >0$ define a 3D pentahedron in the six-dimensional moduli space spanned
by $1/g^2_1,...,1/g^2_6$. See details in \cite{Goran}.
The interactions Lagrangian's in terms of physical fields and effective couplings  are \cite{Goran}:
\begin{eqnarray}
{\cal L}^{\theta}_{\gamma\gamma\gamma}&=&\frac{e}{4} \sin2{\theta_W}\;
{{\rm K}_{\gamma\gamma\gamma}}
h\theta^{\rho\tau}A^{\mu\nu}\left({ a}A_{\mu\nu}A_{\rho\tau}-4A_{\mu\rho}A_{\nu\tau}\right)\,,
\label{22}\\
& & \nonumber \\
{\cal L}^{\theta}_{Z\gamma\gamma}&=&\frac{e}{4} \sin2{\theta_W}\,
{{\rm K}_{Z\gamma \gamma}}\,
h\theta^{\rho\tau}
\left[2Z^{\mu\nu}\left(2A_{\mu\rho}A_{\nu\tau}-{ a}A_{\mu\nu}A_{\rho\tau}\right)
\right.\nonumber\\
& & +\left. 
8 Z_{\mu\rho}A^{\mu\nu}A_{\nu\tau} - 
{ a}Z_{\rho\tau}A_{\mu\nu}A^{\mu\nu}\right]\,,\;\;\;{\rm ect.},
\label{23} \\
{\rm K}_{\gamma\gamma\gamma}&=&\frac{1}{2}\; gg'(\kappa_1 + 3 \kappa_2)\,,
\;\;\;\;\;
{\rm K}_{Z\gamma\gamma}=\frac{1}{2}\; 
\left[{g'}^2\kappa_1 + \left({g'}^2-2g^2\right)\kappa_2\right]\,,\;\;\;{\rm ect.} 
\label{24}
\end{eqnarray} 

\section{Renormalization}
One-loop renormalization is performed by using 
the background field method (BFM) \cite{'tHooft:1973us,PS}.
Advantage of the BFM is the guarantee
of covariance, because by doing the path integral the local symmetry
of the quantum field ${\Phi}_V$ is fixed, while the gauge
symmetry of the background field ${\phi}_V$ is manifestly preserved.
Quantization is performed by the
functional integration over the quantum vector field ${\bf
\Phi}_V$ in the saddle-point approximation around classical
(background) configuration. For case $\phi _V= constant$,
the main contribution to the
functional integral is given by the Gaussian integral.
Split the vector potential
into the classical background plus the quantum-fluctuation parts,
that is: We replace, $\phi_V\to \phi_V + {\bf\Phi}_V$, and than compute the
terms quadratic in the quantum fields. 
Interactions are of the polynomial type.

Proper quantization requires the presence
of the gauge fixing term $ {S_{\rm gf}[\phi]}$. 
Adding to the SM part in the usual way,
Feynman-Faddeev-Popov ghost appears in the effective action. Result of functional integration 
\begin{eqnarray}
\Gamma [\phi] &=& S_{\rm cl}[\phi] + S_{\rm gf}[\phi] + \Gamma^{(1)}[\phi]\,,
\;\;\;\;\;
 S_{\rm gf}[\phi]=-\frac 12 \int
\mathrm{d}^4x(D_{\mu}\bf\Phi_V^{\mu})^2\;, 
\label{25}
\end{eqnarray}
produce the standard result of the commutative part of our action.
The one-loop effective part $\Gamma^{(1)} [\phi]$ is given by
\begin{equation}
\Gamma ^{(1)}[\phi] =\frac{\mathrm{i}}{2}\log\det
S^{(2)}[\phi]=\frac{\mathrm{i}}{2} {\rm Tr}\log S^{(2)}[\phi]\,,
\label{26}
\end{equation}
where $S^{(2)}[\phi]$ is the second functional derivative of a classical action.

The one-loop effective action computed by using background field method gives
\begin{eqnarray}
\Gamma^{(1)}_{\theta,2}&=& \frac{\mathrm{i}}{2} {\rm Tr} \log
\left(\mathcal I + \square^{-1} (N_1+N_2+T_1+T_2+T_3+T_4)\right)
\label{27}\\
&=& \frac{\mathrm{i}}{2}\sum_{n=1}^\infty \frac{(-1)^{n+1}}{n} {\rm Tr}
\left(\square^{-1}N_1+\square^{-1}N_2+\square^{-1}T_1
+\square^{-1}T_2 +\square^{-1}T_3+\square^{-1}T_4\right)^n\,, 
\nonumber
\end{eqnarray}
where $N_i$ are commutative and $T_i$ noncommutative vertices, respectively
\cite{Buric:2005xe,Buric:2006wm,Latas:2007eu,Buric:2007ix,Buric:2004ms}.

\subsection{\bf Renormalization of nmNCSM}
Divergent contributions for this model comes from combinations of $N_1,\;N_2$ and $T_1,\;T_2$ vertices.
Divergences for $\rm U(1)_Y-SU(2)_C$ and $\rm U(1)_Y-SU(3)_C$ mixed 
noncommutative terms, from (\ref{21}), are
\begin{eqnarray}
\Gamma^{(1)}_{\rm div}&=& \frac {11}{3 (4\pi)^2\epsilon}\int d^4 x B_{\mu\nu}^iB^{\mu\nu i} +
\frac {11}{2 (4\pi)^2\epsilon}\int d^4 x G_{\mu\nu}^aG^{\mu\nu a}
\nonumber \\ 
&+& 
\frac{4}{3(4\pi)^2\epsilon}g^\prime g^2  \kappa_2(3-{ a})h\theta^{\mu\nu}\int d^4 x  
\big(\frac 14 f_{\mu\nu}B_{\rho\sigma }^i B^{\rho\sigma i} - f_{\mu\rho}B_{\nu\sigma }^i B^{\rho\sigma i}
 \big)
 \nonumber \\
&+& \frac{6}{3(4\pi)^2\epsilon}g^\prime g^2_S \kappa_3(3-{ a})h\theta^{\mu\nu}\int d^4 x  
\big(\frac 14 f_{\mu\nu}G_{\rho\sigma }^a G^{\rho\sigma a} - 
f_{\mu\rho}G_{\nu\sigma }^a G^{\rho\sigma a}
 \big)\,.
 \label{28}
 \end{eqnarray}
Renormalization is obtained via counter-terms  and for the obvious choice $a=3$,
giving bare Lagrangian
 \begin{eqnarray}
 {\cal L} +{\cal L}_{ct}&=&
-\frac{1}{4}{f_0}_{ \mu\nu }{f_0}^{\mu\nu }-\frac{1}{4}{B_0}_{ \mu\nu
}^i{B_0}^{\mu\nu i} -\frac{1}{4}{G_0}_{ \mu\nu }^a{G_0}^{\mu\nu a}
\nonumber\\
&+&
g^{\prime 3}\kappa_1 h\theta^{\mu\nu} \left( \frac 34 {f_0}_{\mu\nu}{f_0}_{\rho\sigma}{f_0}^{\rho\sigma}-
{f_0}_{\mu\rho}{f_0}_{\nu\sigma}{f_0}^{\rho\sigma}\right)
\nonumber\\
&+&
g_0^\prime g_0^2  \kappa_2 h\theta^{\mu\nu}
\left(\frac 34 {f_0}_{\mu\nu}{B_0}_{\rho\sigma }^i B_0^{\rho\sigma i}
-{f_0}_{\mu\rho}{B_0}_{\nu\sigma }^i B_0^{\rho\sigma i}+c.p.\right)
 \nonumber \\
&+&
g_0^\prime (g_{S})_0^2  \kappa_3 h\theta^{\mu\nu}
\left(\frac 34 {f_0}_{\mu\nu}{G_0}_{\rho\sigma }^a G_0^{\rho\sigma a}-{f_0}_{\mu\rho}{G_0}_{\nu\sigma 
}^a G_0^{\rho\sigma a}+c.p.\right)\,.
\label{29}
 \end{eqnarray}
In the above expression the bare quantities are:
 \begin{eqnarray}
{\cal A}_0^{\mu}&=&{\cal A}^{\mu }\, ,\qquad g_0 ^\prime = g ^\prime\, ,
\nonumber\\
{B_0}^{\mu i}&=&B^{\mu i}\sqrt{1+\frac{44g^2}{3(4\pi)^2\epsilon}}\, ,\quad g_0 =\frac{g\,\mu^{\epsilon/2}}
{\sqrt{1+\frac{44g^2}{3(4\pi)^2\epsilon}}}\, ,
\nonumber\\
{G_0}^{\mu a}&=&G^{\mu a}\sqrt{1+\frac{22g_S^2}{(4\pi)^2\epsilon}}\,,\quad{(g_{S})}_0=\frac{g_S\,\mu^{\epsilon/2}}
{\sqrt{1+\frac{22g_S^2}{(4\pi)^2\epsilon}}}\,  .
\label{30}
 \end{eqnarray}
Constants $\kappa_1$, $\kappa_2$ and $\kappa_3$ remain  
unchanged under renormalization
\begin{eqnarray}
 \kappa_1=\,{(\kappa_{1})}_0\,, \:\:\;\kappa_2=\,{(\kappa_{2})}_0\,,
 \:\:\;\kappa_3=\,{(\kappa_{3})}_0\,,
 \label{31}
 \end{eqnarray}
if the following replacement in $1/g^2_i$ couplings were applied:
  \begin{eqnarray}
 \frac{1}{g_1^2}&=&{(\frac{1}{g_1^2})}_0+\frac{33}{18(4\pi)^2\epsilon}\,, 
 \hspace{.5cm}
 \frac{1}{g_2^2}= {(\frac{1}{g_2^2})}_0+\frac{-11}{18(4\pi)^2\epsilon}\,,
 \hspace{.5cm}
 \frac{1}{g_3^2}={(\frac{1}{g_3^2})}_0+\frac{-11}{18(4\pi)^2\epsilon}\,,
 \nonumber\\
 \frac{1}{g_4^2}&=&{(\frac{1}{g_4^2})}_0+\frac{-143}{18(4\pi)^2\epsilon}\,,
 \hspace{.5cm}
 \frac{1}{g_5^2}={(\frac{1}{g_5^2})}_0+\frac{-121}{18(4\pi)^2\epsilon}\,,
 \hspace{.5cm}
 \frac{1}{g_6^2}={(\frac{1}{g_6^2})}_0+\frac{110}{18(4\pi)^2\epsilon}\,.
 \label{32}
 \end{eqnarray}
Since, for $a=3$, our Lagrangian is free from divergences at one-loop
noncommutative deformation parameter $h$ need not be renormalized.

\subsection{\bf Renormalization of noncommutative SU(N) gauge theory and mNCSM gauge sector}
Choosing vector field in the adjoint representation SU(N) we have
the following Lagrangian
\begin{eqnarray}
S_{cl}&=&S_{\mathrm{NCYM}}=\int \mathrm{d}^4x \left(-\frac{1}{4}
F^a_{\mu\nu} F^{a\mu\nu}
+ \frac{1}{4} h\theta^{\mu\nu} d^{abc}
\left( \frac{ a}{4} F^a_{\mu\nu} F^b_{\rho\sigma} - F^a_{\mu\rho}
F^b_{\nu\sigma} \right) F^{c\rho\sigma} \right),
 \label{33}
\end{eqnarray}
where $d^{abc}$ are totally symmetric coefficients of the SU(N) group which come from the trace
in (\ref{15}); $a,b,c=1,...,N^2-1$ are the group indices.
Divergent contributions for the model, computed by using BFM, comes from combinations of 
$N_1,\;N_2$ and $T_2,\;T_3,\;T_4$ vertices.
Renormalization of the theory is obtained by canceling  divergences.
To have that the counter terms should be
added to the starting action, which than produces the bare Lagrangian
\begin{eqnarray}
\mathcal{L}_0&=&-\frac{1}{4} {F_0}^a_{\mu\nu}{F_0}^{a\mu\nu} +\frac{1}{4}
g\mu^{\epsilon/2} h \theta^{\mu\nu}d^{abc} 
\nonumber\\
&\times&
\left[\frac{ a}{4}
\left(1-\frac{3-25{ a}}{3{ a}}\frac{Ng^2}{(4\pi)^2\epsilon}\right)F^a_{\mu\nu}
F^b_{\rho\sigma}
-\left(1+\frac{21+{ a}}{3}\frac{Ng^2}{(4\pi)^2\epsilon}\right)F^a_{\mu\rho}
F^b_{\nu\sigma} \right]F^{c\rho\sigma}\,.
\label{34}
\end{eqnarray}
To reach the same structure as in starting Lagrangian we have to impose the condition
\begin{equation}
\left(-\frac{25{ a}-3 }{48}\right):\left(\frac{{ a}+21}{12}\right) =
\frac{{ a}}{4} : (-1), 
\label{35}
\end{equation}
which has two solutions: $a=1$ and $a=3$ \cite{Latas:2007eu}.

The case $a=1$ corresponds to previous result \cite{Buric:2005xe} and  
the deformation parameter ${h}$ 
need not to be renormalized. Renormalizability is, in this case, obtained through the known renormalization 
of gauge fields and coupling constant only.

However the case $a=3$ is different since additional divergences can be absorbed
only into the noncommutative deformation parameter ${h}$. That is that ${h}$ has
to be renormalized. The bare gauge field, the coupling constant and 
the noncommutative deformation parameter are \cite{Latas:2007eu}:
\begin{eqnarray}
V^\mu_0&=&V^\mu\sqrt{1+\frac{22Ng^2}{3(4\pi)^2
\epsilon}},
\;\;\;\;\;
g_0=\frac{g\mu^{\epsilon/2}}{\sqrt{1+\frac{22Ng^2}{3(4\pi)^2
\epsilon}}},
\;\;\;\;\;
h_0=\frac{h}{1-\frac{2Ng^2}{3(4\pi)^2\epsilon}}\,. 
\label{36}
\end{eqnarray}
The necessity of the $ {h}$ renormalization jeopardizes previous hope
that the {NC SU(N)} gauge theory might be renormalizable to all orders
in $ {\theta^{\mu\nu}}$. Above results are also valid for 
the minimal NCSM gauge sector (\ref{18}) with $N=3$. 

\subsection{\bf Ultraviolet asymptotic behavior of noncommutative SU(N) gauge theory}

Gauge coupling constant ${g}$ in our theory 
depends on energy i.e., the renormalization point ${\mu}$, {\it satisfying the same
beta function} as in QCD
\begin{equation}
\beta_g=\mu \frac{\partial }{\partial\mu}g(\mu)
=-\frac{11Ng^3(\mu)}{3(4\pi)^2}
\;\;\;\;\Rightarrow\;\;\;\;
\alpha_s(\mu)=\frac{g^2(\mu)}{4\pi}=
\frac{6\pi}{11N}\frac{1}{\ln\frac{\mu}{\Lambda}}\,, 
\label{37}
\end{equation}
{\it our theory is {\rm UV} stable, i.e. ${asymptotically\; free}$}. 
In (\ref{37}) ${\Lambda}$ is an integration constant 
determined from experiment: 
hadronic production in ${e^+e^-}$ annihilation at the ${Z}$
resonance has given ${\alpha_s(m_Z)=0.12}$
corresponding to ${\Lambda= \Lambda_{\rm QCD}\simeq 250}$ MeV.
Next, from (\ref{36}) and (\ref{37}) we have
\begin{equation}
\beta_h=\mu \frac{\partial}{\partial\mu}\,h(\mu)
=-\frac{11Ng^2(\mu)}{24\pi^2}\,h(\mu)
\;\;\;\;\Rightarrow\;\;\;\;
h(\mu) = \frac{h_0}{\ln\frac{\mu}{\Lambda}}\,. 
\label{38}
\end{equation}
Both ${\beta}$ functions are {\it negative} that is it
decrease with increasing energy ${\mu}$ \cite{Latas:2007eu}. 
Solution to ${\beta_h}$ shows that by increase of energy ${\mu}$ the the NC deformation
parameter ${h}$ decreases. The NC deformation parameter ${h}$ 
becomes {\it the running deformation parameter
and vanishes for large ${\mu}$} \cite{Latas:2007eu}. From this follows necessity of the 
modification of Heisenberg uncertainty relations at high energy. 
String theory inspired modification
\begin{eqnarray}
[x,p]&=&i\hbar(1+\beta p^2)
\;\;\;\;\Rightarrow\;\;\;\;
\Delta x=\frac{\hbar}{2}(\frac{1}{\Delta p}+\beta \Delta p).
\label{39}
\end{eqnarray}
show that for large momenta ${\Delta p}$ (energy) distance ${\Delta x}$ grows linearly.  
So large energies do not necessarily correspond to small distances, and running ${h}$ does
not imply that noncommutativity vanishes at small distances.
This is related to {UV/IR} correspondence. From Eq. (\ref{38}) and $h=1/\Lambda_{\rm NC}$
we have
\begin{equation}
h(\mu)=\frac{1}{\Lambda^2_{\rm NC}(\mu)}
\;\;\;\;\Rightarrow\;\;\;\;
\Lambda_{\rm NC}(\mu) =\Lambda_{\rm NC}\,\sqrt{\ln
\frac{\mu}{\Lambda}}\,, 
\label{40}
\end{equation}
i.e. ${\Lambda_{\rm NC}}$ becomes a function of energy ${\mu}$.
This way, via RGE, the scale of noncommutativity ${\Lambda_{\rm NC}}$ 
becomes the running scale of non-commutativity too \cite{Latas:2007eu}. 
However it receives very is small change when energy ${\mu}$ increases.
This means that there is a large degree of 
stability of NC SU(N) theory within a wide range of energies.
For example, considering typical QCD energies, ${\mu=m_Z}$, factor
 ${\sqrt{\ln({m_Z}/{\Lambda_{\rm QCD}})}\simeq 2.4}$.

\subsection{\bf The 4$\psi$ divergences for noncommutative chiral fermions in U(1) and SU(2) cases}
The one-loop effective action is computed from Eq. (\ref{16}) by using 
background field method
\begin{eqnarray}
\Gamma^{(1)}_{\theta,2}&=& \frac{\mathrm{i}}{2} \mathrm{STr} \log
\left(\mathcal I + \square^{-1} (N_1+N_2+T_1+T_2+T_3)\right)
\nonumber\\
&=& 
\frac{\mathrm{i}}{2}\sum_{n=1}^\infty \frac{(-1)^{n+1}}{n} \mathrm{STr}
\left(\square^{-1}N_1+\square^{-1}T_1+\square^{-1}T_2 \right)^n\,. 
\label{41}
\end{eqnarray}
Divergent contributions comes from 
\begin{eqnarray}
\mathcal{D}_1 =\mathrm{STr}\left((\square ^{-1}N_1)^3 (\square ^{-1}T_1)
\right)~,\;\;
\mathcal{D}_2 =\mathrm{STr} \left((\square ^{-1}N_1)^2 (\square ^{-1}T_2)
\right)~.
\label{42}
\end{eqnarray}
Our computations shows that term ${\cal D}_1$ 
is finite due to the structure of the momentum integrals in both, the U(1)
and the SU(2), cases.
For NC chiral electrodynamics, U(1), with Majorana spinors 
and with the usual definition for the supertrace $\mathrm{STr}$, \cite{Buric:2004ms},
we have found
\begin{equation}
\mathcal{D}^{\rm U(1)}_2\vert_{\mathrm{div}} =\frac{1}{(4\pi )^2\epsilon}
\frac{3\mathrm{i}}{8}h\theta^{\mu\nu}\varepsilon_{\mu\nu\rho\sigma}
(\bar\psi\gamma^\rho\gamma_5\psi)
(\bar\psi\gamma^\sigma\gamma_5\psi)\equiv 0\,.
\label{43}
\end{equation}

Next we have used chiral fermions in the fundamental representation of SU(2). 
Choosing Majorana spinors we apparently break the SU(2) symmetry, and consequently  
we have to work in the framework of the components for the vector potential.
Now of course, Majorana $\begin{pmatrix}\psi_1 \\ \psi_2 \end{pmatrix} $ is not a SU(2) doublet.
Divergent part of $\mathcal{D}_2$ is 
\begin{equation}
\mathcal{D}^{\rm SU(2)}_2\vert_{\mathrm{div}} = \frac{-1}{(4\pi)^2\epsilon}\,
\frac{9\mathrm{i}}{64} h\theta^{\mu\nu} \varepsilon_{\mu\nu\rho\sigma}
(\bar\psi_1\gamma^\rho\gamma_5\psi_1
+\bar\psi_2\gamma^\rho\gamma_5\psi_2)
(\bar\psi_1\gamma^\sigma \gamma_5\psi_1 +
\bar\psi_2\gamma^\sigma \gamma_5\psi_2),
\label{44}
\end{equation}
and it vanishes identically, too. 

Clearly, we may conclude that direct computations by using BFM confirms results of 
the symmetry analysis for the 4$\psi$ divergent term, 
which, due to its SU(2) invariance, has to be zero \cite{Buric:2007ix}.
The same symmetry arguments holds also for U(1) and SU(2) $\mathcal{D}_1$ terms,
i.e. they both vanish identically too.

\section{Forbidden decays $Z \to \gamma\gamma,\;gg$}
From  the gauge-invariant amplitude for $Z \to \gamma\gamma,\;gg$ decays in momentum space
and for $Z$ boson at rest we have found the following branching ratios.
For $a=3$, we have
\begin{eqnarray}
BR(Z\rightarrow \gamma\gamma)
&=&\tau_Z \frac{\alpha}{4} \;\frac{M^5_Z}{\Lambda^4_{\rm NC}}\; \sin^2 2\theta_W
{\rm K}^2_{Z\gamma \gamma}
\big({\vec E}_{\theta}^2 + {\vec B}_{\theta}^2\big)
=\frac{1}{8}\frac{{\rm K}^2_{Z\gamma \gamma}}{{\rm K}^2_{Zgg}}BR(Z\rightarrow gg)\,,
\label{45}
\end{eqnarray}
where $\tau_Z $ is the $Z$ boson lifetime. 
LHC experimental possibilities for $Z \to \gamma\gamma$ 
we analyze by using the CMS Physics Technical Design Report \cite{CMS1,CMS2}.
We have found that for $10^7$ events of $Z\rightarrow e^+e^-$ for $10\;fb^{-1}$ in 2 years of LHC
running and by assuming $BR(Z\rightarrow \gamma\gamma) \sim 10^{-8}$ and 
using $BR(Z\rightarrow e^+e^-) = 0.03 $ about $ \sim 3$ events
of $Z\rightarrow \gamma\gamma$ decays should be found. 
However, note that background sources (CMS Note 2006/112, Fig.3) could potentially be a big problem.
For example study for $Higgs\rightarrow \gamma\gamma$ shows that, 
when ${e^-}$ from $Z\rightarrow e^+e^-$  radiates 
very high energy Bremsstrahlung photon
into pixel detector, for similar energies of 
${e^-}$ and ${\gamma}$, there is a
huge probability of misidentification of 
${e^-}$ with ${\gamma}$. Second, the irreducible di-photon background
may also kill the signal. The $Z \to gg$ decay was discussed in \cite{Goran}.

Finally, note that after 10 years of LHC running integrated 
luminosity would reach $\sim  1000\;fb^{-1}$.
In that case and from bona fide reasonable assumption $BR(Z\rightarrow \gamma\gamma) \sim 10^{-8}$ 
one would find ~$300$ events of $Z\rightarrow \gamma\gamma$
decays, or one would have $ \sim 3$ events with 
$BR(Z\rightarrow \gamma\gamma) \sim 10^{-10}$.
From above it follows that, in the later case, the lower bound on the
scale of noncommutativity would be $\Lambda_{\rm NC}\,>\, 1\; {\rm TeV}$.

\section{Limits on the noncommutativity scale $\Lambda_{\rm NC}$}
Limits on the scale of noncommutativity in high energy particle physics 
are coming from the analysis of decay and scattering experiments. 

Considering SM forbidden decays, recently we have found 
the following lower limit $\Lambda_{\rm NC} > 1$ TeV \cite{Buric:2007qx}
from $Z\rightarrow \gamma\gamma$ decay.
Note here that earlier limits obtained from 
 $\gamma_{\rm pl}\to \nu \bar\nu$ decay (astrophysics analysis) produces
$\Lambda_{\rm NC} > 81$ GeV \cite{Josip}
while from the SM forbidden
$J/\psi\to\gamma\gamma$ and $K\to\pi\gamma$ 
\cite{Melic:2005hb} 
decays we obtain  
$\Lambda_{\rm NC} > 9$ GeV, and
$\Lambda_{\rm NC} > 43$ GeV, respectively. 
Last two bounds are not usefull 
due to the too high lower limit of the relevant branching ratios.

Scattering experiments \cite{Ohl:2004tn} support the above obtained limits.
From annihilation $\gamma\gamma\to\,{\bar f}f$ it was found $\Lambda_{\rm NC} > 200$ GeV,
which is a bit to low. However, 
from ${\bar f}f \to Z\gamma $ unelastic scattering experiments
there is very interesting limit $\Lambda_{\rm NC} > 1$ TeV.

\section{Summary and Conclusion}
Principle of renormalizability implemented on 
our {$\theta$}-expanded NCGFT led us to well defined deformation via introduction of
higher order noncommutative action class for the gauge sectors of 
the mNCSM, nmNCSM and NC SU(N) models.
This extension was parametrized by generically free parameter $a$:
\begin{equation}
 S_g =-\frac{1}{2}{\rm Tr}\int d^4x
\left(1
+\mathrm{i}({a}-1)\,\widehat x_{\rho}\star \widehat x_{\sigma}
\star\widehat F^{\rho\sigma}\right)\star\widehat F_{\mu\nu}
\star\widehat F^{\mu\nu}\,.
\label{46}
\end{equation}
\normalsize
We have found the following properties of the above models 
with respect to renormalization procedure:
\\
$\star$ Renormalization principle is fixing the freedom parameter $a$
for our $ \theta$-expanded NC GFT.
\\
$\star$ Divergences cancel differently than in 
commutative GFT and this depends on the representations.
\\
$\star$ Gauge sector of the mNCSM is renormalizable for $ {a=1}$. 
Divergences were absorbed through the coupling and fields redefinition only,
like in the SM. Consequently, no renormalization of the 
noncommutative deformation parameter ${h}$ is necessary.
\\
$\star$ Gauge sector of the nmNCSM, which produces SM forbidden $Z\to \gamma\gamma$ decay, is
renormalizable and {\it \bf{finite}} for ${a=3}$. 
Due to this finiteness no renormalization of ${h}$ necessary.
\\
$\star$ Noncommutative SU(N) gauge theory is renormalizable for ${a=1}$ and ${a=3}$. 
The case ${a=1}$ corresponds to the earlier obtained result \cite{Buric:2005xe}. 
However, in the case ${a=3}$
additional divergences appears and had to be absorbed through the renormalization
of the noncommutative deformation ${h}$. Hence, in the case of noncommutative SU(N) 
the noncommutativity deformation parameter 
$ {h}$  had to be renormalized and it is 
{\it \bf{asymptotically free}}, opposite to the previous expectations.  
\\
$\star$ The solution ${a=3}$, while shifting the model
to the higher order, i.e. while extending `NC minimal coupling', 
hints into the discovery of  
the key role of the higher noncommutative gauge interaction in
one-loop renormalizability of classes of NCGFT at the first order in $\theta$. 
\\
$\star$ Our computations also confirms symmetry arguments that 
for noncommutative chiral electrodynamics, 
that is the U(1) case with Majorana spinors, 
the $4\psi$ divergent part vanishes.
For noncommutative chiral fermions in the fundamental representation of SU(2)
with Majorana spinors the $4\psi$ divergent part vanishes due to the SU(2) invariance.
So, for noncommutative U(1) and SU(2) chiral fermion models 
typical $4\psi$-divergence is {\it \bf{absent}}, contrary to the earlier results obtained  
for Dirac fermions \cite{Wulkenhaar:2001sq,Buric:2004ms}.
\\
$\star$ There is similarity to noncommutative $\phi^4$ theory. Namely, by adding
$\Omega \,\int d^4x\,\widehat x \star \widehat x 
\star \widehat \phi \star \widehat \phi$ term to the `minimal'
action, the noncommutative $\phi^4$ theory becomes renormalizable 
up to all orders \cite{Grosse:2005da}. This way
renormalization principle determines
noncommutative renormalizable deformation up to all orders.
\\
$\star$ Note also that the
renormalizability principle could help to minimize or even cancel
most of the ambiguities of the higher order Seiberg-Witten
maps~\cite{Moller:2004qq}.
\\
$\star$ Finally, phenomenological results, as the standard model forbidden
$Z\to \gamma\gamma$ decay, are 
{\it \bf{robust}} due to the one-loop renormalizability and
{\it \bf{finiteness}} of the nmNCSM gauge sector \cite{Buric:2006wm,Buric:2007qx}.

\begin{acknowledgement}
Part of this work was done during my visit to ESI, Vienna and MPI, M\" unchen.
I would like to use this opportunity to acknowledge 
H. Grosse at ESI, and W. Hollik at MPI, for hospitality and support.
This work is supported by the project
098-0982930-2900 of the Croatian Ministry of Science Education and Sports
and by the European Community's Marie-Curie Research Training Network under
contract MRTN-CT-2006-035505 `Tools and PrecisionCalculations for Physics
Discoveries at Colliders' (HEPTOOLS).

\end{acknowledgement}

\end{document}